\documentclass[times, 10pt,twocolumn]{article}
\usepackage{latex8}
\usepackage{times}
\usepackage{url}
\usepackage{comment}
\usepackage{epsfig}
\usepackage{fancyvrb}

\newcommand{\qed}{$\Box$}

\begin{document}

\advance\baselineskip by -1.2pt

\title{Fast Query Processing by Distributing an Index over CPU Caches}

\author{Xiaoqin Ma \thanks{This work was partially supported by
                the National Science Foundation under Grants
                CCR-0204113 and ACIR-0342555, and
                by the
                Institute for Complex Scientific Software
                (ICSS, http://www.icss.neu.edu/).}
 and Gene Cooperman$^*$ \\
 College of Computer and Information Science, \\
 and Institute for Complex Scientific Software \\
 Northeastern University \\
 Boston, MA 02115 USA \\
 \{xqma, gene\}@ccs.neu.edu
}

\maketitle
\thispagestyle{empty}

\begin{abstract} Data intensive applications on clusters often require
 requests quickly be sent to the node managing the desired data.  In
 many applications, one must look through a sorted tree structure to
 determine the responsible node for accessing or storing the data.
 Examples include object tracking in sensor networks, packet routing
 over the internet, request processing in publish-subscribe
 middleware, and query processing in database systems.  When the tree
 structure is larger than the CPU cache, the standard implementation
 potentially incurs many cache misses for each lookup; one cache miss
 at each successive level of the tree.  As the CPU-RAM gap grows, this
 performance degradation will only become worse in the future.

  We propose a solution that takes advantage of the growing speed of
  local area networks for clusters.  We split the sorted tree
  structure among the nodes of the cluster. We assume that the
  structure will fit inside the aggregation of the CPU caches of the
  entire cluster.  We then send a word over the network (as part of a
  larger packet containing other words) in order to examine the tree
  structure in another node's CPU cache.  We show that this is often
  faster than the standard solution, which locally incurs multiple
  cache misses while accessing each successive level of the tree.
  
  The principle is demonstrated with a cluster configured with Pentium
  III nodes connected with a Myrinet network.  The new approach is
  shown to be 50\% faster on this current cluster.  In the future, the
  new approach is expected to have a still greater advantage as
  networks grow in speed, and as cache lines grow in length (greater
  cache miss penalty).  This can be used to successfully overcome the
  inherent memory latency associated with cache misses.
\end{abstract}

\section{Introduction}

In the past decade, hardware technology had two trends. On the one
hand, microprocessor speeds have followed Moore's Law~\cite{Moor65},
doubling every eighteen months. In the future, we may instead see a
doubling of the number of processors per chip, such as
multicore/multiprocessor chips, but the effect on computational power
is the same.  However, the development of memory shows a different
trend. Although memory capacity and prices are keeping up with the
increase rate of CPU speed, memory latency has improved little.
Memory latency for random access to a RAM chip runs into a fundamental
lower bound determined by the time to precharge the chip buffer from
internal voltages to the voltages needed to drive an external bus.
Hence, newer memory standards, such as DDR2 RAM and Rambus RAM,
concentrate on improving memory bandwidth, but not memory
latency. Over the past decade, the gap between CPU speed and memory
latency has increased exponentially.  Current technology trends
(especially the higher memory pressures due to the introduction of
dual and quad processor CPU cores) portend further increases in this
gap.

This CPU-memory gap represents a fundamental bottleneck in distributed
applications that require messages or queries to quickly be routed to
appropriate nodes, based on a large index data structure.  Two
examples of such an index data structure are a sorted n-ary tree and a
sorted array.  In the case of the sorted array, one can look up an
index via binary search.

We do not consider hash arrays for the index data structure.
Specifically, we assume that a key is part of a very large ordered
index set.  The range of all possible indices is divided into
sub-ranges.  For example, if the indices have values from 0.0 to~1.0,
then there might be three nodes in charge of indices from 0.0 to~0.33,
from 0.33 to~0.67, and from 0.67 to~1.0, respectively.  In this simple
example, the index data structure would record the delimiters 0.0,
0.33, 0.67 and~1.0.  The key specified by an incoming query could be
any number between 0.0 and~1.0.

We also assume that multiple nodes are available for passing external
queries to the correct node, based on the key value in the incoming
query.  The key must be looked up in the index.  We further assume
that the index is too large to fit in the CPU cache, and overflows
into main RAM.  Examples include tracing objects in sensor networks,
routing packets over internet, routing requests in publish-subscribe
middleware, and query processing with database indices.

In this situation, rather than replicate the index on each node, we
propose to distribute the index among the CPU caches of the multiple
nodes.  We assume that the aggregate CPU cache of the multiple nodes
is sufficient to hold the index.  We consider three variations of this
idea.

We compare each of them with two standard methods (here called
Method~A and Method~B).  Methods~A and~B each duplicate the index
structure on each node, accept queries at a single dispatcher node
which dispatches queries to an appropriate node according to a load
balancing algorithm, and each other nodes lookup the duplicated index
structure in memory and dispatch the results to the target.  The three
variations of Method~C have only one copy of the index structure among
all the nodes, accept queries on a single {\em master node}.  The
master node passes the query to an appropriate {\em slave node}
according to one piece of index structure stored on it, then each
slave node processes the queries over one piece of index stored on it
and dispatches the results to the target.  For all methods, the $n$ in
an n-ary tree is chosen so that $n$~keys ($n$~4-byte words in our
case) and the corresponding pointers fit exactly in an L2 cache line.
\begin{itemize}
\item{\bf Method A} --- index is a large n-ary tree and is duplicated
               on each node; at each node, each query incurs multiple
               cache misses.
\item{\bf Method B} --- index is a large n-ary tree and is duplicated
                on each node; at each node, many queries are stored
                and then processed as a batch; to process a batch of
                queries, a single pass through the tree is made with
                a buffering access technique using the L2~cache (see
                Section~\ref{sect:methodBdesc}).
\item{\bf Method C} --- index is a large sorted array and is
                partitioned among the nodes; with each slave node
                holding one partition.  The master node holds the
                delimiters for the partitions.
\begin{itemize}
\item{\bf Method C-1} --- the partition on the slave node is stored as
an n-ary tree.
\item{\bf Method C-2} --- the partition on the slave node is stored as
                an n-ary tree; As with Method~B, queries are stored
                and processed in a batch. To process a batch of
                queries, a single pass through the tree is made with
                the buffering access technique, but using the L1~cache
                instead of the L2~cache (see Section~\ref{sect:methodCdesc}).
\item{\bf Method C-3} --- the partition on the slave node is stored as
a sorted array.
\end{itemize}
\end{itemize}

Method~C is the novel method of {\em distributed in-cache index}
(based on aggregating the CPU cache from multiple nodes).  The
distributed in-cache index is formally defined in
Section~\ref{sect:definitions}, and contrasted to traditional {\em
cooperative caching} (based on aggregating the RAM from multiple
nodes).  Method~B is based on the buffering access technique,
described by Zhou and Ross~\cite{Zhou03}.
Section~\ref{sect:algorithms} describes all of the methods studied
here.  In the experimental section (Section~\ref{sect:comparison}), we
demonstrate that Method~C-3 is the best for simultaneously satisfying
the two criteria of throughput and response time.

\smallskip
\noindent
{\bf Modeling the Future.}  Although Method~C-3 is somewhat faster
today, it is important to demonstrate that the advantage of Method~C-3
will widen further in the future.  This is important as CPU speed, memory
bandwidth, and network speed all increase.  In order to predict the
speeds of the five methods using future technology, we first define a
simple analytical model that successfully analyzes the running time of
the five methods on today's architecture.  Our analytical models are
based on architectural parameters of the technology employed.

The analytical model was first checked for accuracy against the
Methods~A, B and~C-3.  (Methods~C-1 and C-2 could also be analyzed,
but current experiments showed them to be inferior to C-3.)  The
analytical model was found to be accurate within 25\% for the three
methods analyzed.

We then make reasonable assumptions about technology trends, in order
to plug in architectural parameters for future technologies.
Appendix~\ref{sect:analysis} describes the analytical model that
predicts the performance of the three methods.
Section~\ref{sect:expmodel} demonstrates future trends of the three
methods based on the model.

\subsection{Related Work}
\label{sect:related}
The concept of the memory wall has been popularized by
Wulf~\cite{WulfMcKee95}.  Many researchers have been working on
improving cache efficiency to overcome the memory wall problem. The
pioneering work~\cite{LamETAl91} done by Lam et al. has both
theoretically and experimentally studied the blocking technique and
described the factors that affect the cache formance. However, there
is not an easy way to apply the blocking technique to the tree traversal
problem or to the index structure lookup problem to improve the cache
efficiency.

The issue of cache and n-ary trees is closely related to the issue of
memory-resident B+-trees.  There is a large stream of research on this
in the database community ~\cite{ Chen02, Goetz01, Hankins03, Rao00,
Zhou03}.  Rao~\cite{Rao00} proposed the CSB+ tree (cache sensitive B+
tree).  In a CSB+ tree, the branching factor is improved by storing
only the first child pointer at each node.  Other child pointers can
be calculated by adding the offset to the first child pointer because
all child nodes are stored consecutively in the memory space in a CSB+
tree.  Recently, Zhou~\cite{Zhou03} proposed the buffering access
technique to improve the cache performance for a bulk lookup.
However, cache miss penalties still account for over 30\% of the total
cost for each query in all above proposed methods.

In the area of theory and experimental algorithms, Ladner et
al.~\cite{Ladner99} proposed an analytical model to predict the cache
performance.  In their model, they assume all nodes in a tree are
accessed uniformly.  This model is not accurate for the tree lookup
problem. Because the number of nodes from root node to leaf nodes is
exponentially increasing, nodes' access rates are exponentially
decreasing as the their positioned levels in the tree increase.
Hankins and Patel~\cite{Hankins03} proposed a model with an
exponential distributed node access rate in a B+ tree according to the
level of a node positioned.  However, they only considered the
compulsory cache misses, and not the capacity cache misses.  They also
assume that the tree can fit in the cache. So, for tree structures
that can't fit in the cache, the model in~\cite{Hankins03} is not
applicable.

With the development of the technologies, the performance gap between
sequential and random accesses to RAM is increasing due to
difficulties in circuit design, such as the issue of precharging
the buffer.
Cooperman et al.~\cite{CoopermanEtAl04} studied the performance impact
of random accesses to RAM and proposed the MBRAM model that
distinguishes between random and sequential accesses to RAM.  They also
show that tree traversal applications can generate many random
memory accesses resulting in degraded performance, as demonstrated by
heap sort.
In parallel, Byna et al.~\cite{SurEtAl04}
proposed a memory cost model for looping operations.

\section{Distributed in-Cache indices}
\label{sect:definitions}

\subsection{The Definition of Distributed in-Cache Indices}
\label{sect:ourdefinition}

Historically, one often used aggregate memory in a cluster to store
files to reduce the number of disk accesses. 
We explore the use of this technique one level higher in the
memory hierarchy than what is traditionally considered to avoid random
memory accesses.  Because a large
index will not fit in cache, we will partition the index among
the caches of the many nodes in a cluster.  We call this
a {\em distributed in-cache index}.

We design a more effective index lookup strategy over the distributed
in-cache index. The following technology trends stimulate us
to distribute an index over CPU caches in a cluster:
\begin{enumerate}

\item The disparity between processor speed and memory speed is
increasing. As we move to faster, multiple-core CPU chips, the
aggregate processor performance is increasing much more rapidly than
main memory (RAM) performance. This divergence makes it increasingly
important to reduce the number of memory accesses, especially random
memory accesses. Index lookup and tree traversal problems produce many
random memory accesses. For instance, in the Pentium~4, the L2 cache
miss penalty is around~150 ns, which will waste more than~200 CPU
cycles of modern microprocessors.

\item Emerging high-speed low-latency switched networks can transfer
data across the network much faster than standard Ethernet.  The
combined cost of index lookup in the remote L2~cache and data transfer
over an older network might be more expensive than the cost of index
lookup in the local memory. With today's high-speed low-latency
networks, the cost of data transfer in a batch over the network is
lower than the cost of many random accesses to local memory, due to the
stagnating performance of RAM with respect to memory latency in recent
years.  For example, on the Boston University Linux cluster, the
measured random memory bandwidth for a series of 4-byte word accesses
at random locations is
48~MB/s (where each such random access typically incurs a cache miss),
although the sequential memory bandwidth (accessing words in sequence)
is 647~MB/s. The
measured one-way Myrinet bandwidth is 1.1~Gb/s (or 138~MB/s) which is
much faster than the random memory bandwidth.  Further more, in most
of today's systems, communication can overlap with computation. This
makes the communication cost negligible.


\end{enumerate}

\subsection{Design Issues for Distributed in-Cache Indices}
\label{sect:quest}

\paragraph{Network latency:}
Local area network latencies range from the extremely short latency of
Myrinet (approximately 7~$\mu$s) to latencies of about 100~$\mu$s 
for Gigabit Ethernet.  (Further, depending on the protocol stack of the
operating system, the latency seen by the application may be much worse.)
By aggregating many queries into larger,
batched network messages, we can amortize the latency over the
transimission time.  In Myrinet (which is used in our experiments),
the transmision time for a 10~KB message (about 10~KB/(1.1~Gb/s) =
80~$\mu$s) clearly dominates the latency (7~$\mu$s).  For Gigabit
Ethernet, one may need to batch a message as large as 200~KB for the
transmission time to dominate the latency, but the same principle
applies.

\paragraph{Memory bandwidth:}
The memory bandwidth of DDR-266 RAM is 2.1~GB/s, and still faster
variations are available today.  Hence, the full bandwidth of RAM is faster
than the network. 

\paragraph{Memory latency:}
For random memory accesses, memory latency will dominate if
not handled appropriately.  On the Pentium~III, a cache miss for
a 4-byte word will require a 32~byte cache line to be loaded.
Hence, the effective memory bandwidth degrades by at least a factor
of~8.  (In fact, the precharging delay of DRAM technology increases 
the degradation factor.)  The Pentium~4 has a 128~byte cache line, with a
corresponding degradation factor of~32 in the worse case when
successively accessing words are on different cache lines.
(In this random access pattern, each access of a four-byte word requires
loading a new cache line of length 4$\times$32~bytes.)

\paragraph{CPU time:}
We can neglect the CPU time in modeling the overall time for
applications with intensive memory accesses.  This is
because CPU computation and memory access are overlapped, and memory
access time greatly dominates over the time for today's very fast CPUs.

\paragraph{Cache Contention:}
We assume that the aggregate cache size across all CPUs is sufficient
to hold the distributed in-cache index.  As a message of batched
queries is loaded, this will lead to cache pollution by evicting some
portion of the index.  However, the effect of cache pollution is
limited.  For a 4-byte query key, a single cache line of queries will
hold 8~keys on the Pentium~III (and 32~keys on the Pentium~4).
Assuming that query key values are random, each of the 8~queries will
access one leaf node in the index.  Hence, for each cache line of
queries that is processed, we will refresh at least 8~different cache
lines of the tree.  The effect is larger when one considers interior
nodes of the tree.  Further, the Pentium~4 raises this factor from 8
to~32.  Hence, to the extent that a cache eviction algorithm
approximates an LRU algorithm, the probability of evicting a cache
line containing query keys is much larger than the probability of
evicting a cache line containing a part of the index.

\section{Different Index Lookup Methods in a Distributed Environment}
\label{sect:algorithms}

The introduction provided an overview of Methods~A, B and~C.  Method~C
in fact consists of three submethods, C-1, C-2, and~C-3.  Method~A is
a straightforward lookup in a sorted n-ary tree, each node has a
replication of the complete tree. In Method~B, each node also has a
replication of the complete tree, but its
description is more complicated.  We describe Method~B, followed by Method~C.

\subsection{Method~B}
\label{sect:methodBdesc}
\begin{figure}[hbt]
\begin{center}
  \scalebox{0.3}{\epsfig{file=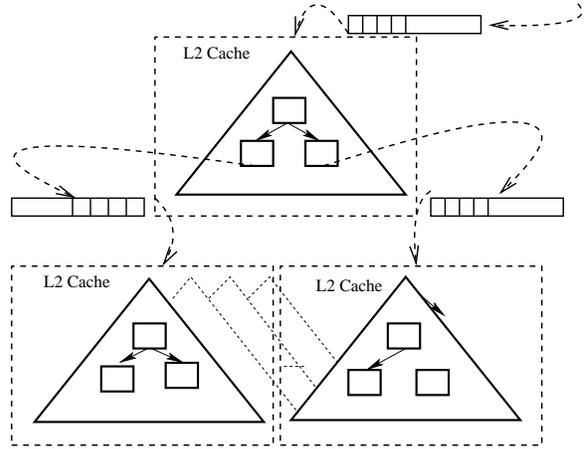}} \caption{Buffering
  Access Method} \label{fig:figB}
\end{center}
\end{figure}

Method~B is based on an idea of Zhou and Ross~\cite{Zhou03}.  They
proposed the buffering access method for a stream of arriving search
keys, as shown in Figure~\ref{fig:figB}.

The index tree is logically decomposed into several subtrees.  A
subtree consists of a root node and all of its descendants, down to
some level~$k$, where $k$ is chosen so that the subtree tree will fit in the
L2~cache.
Along with each subtree, the algorithm maintains an associated buffer
to store search keys that reach the root node of the subtree.

The key to the success of Method~B is to process a batch of search
keys at the same time.  Each key~$k$ in the batch is looked up in the
top level subtree.  The search within the top level subtree will lead
to a leaf node, $x$, of that subtree.  The node~$x$ is also the root
of a lower subtree.  The key~$k$ is then stored into the buffer
associated with the subtree rooted at~$x$.

If there are $\ell$ leaf nodes in the top level subtree, then this
requires streaming write access to $\ell$ buffers.  For $\ell$ of
reasonable size, this process is efficient.
  
After the top level subtree has been processed, each lower subtree is
processed using the keys stored in its buffer as the batch of search keys.
And so the algorithm proceeds recursively.

Since a subtree and its associated buffer can fit inside the L2~cache,
the process is fast, aside from the need to write to different
buffers.  Since the write access is a streaming access, it avoids the
high latency overhead of a cache miss.  Further, such writes can be
non-blocking.

\subsection{Method~C}
\label{sect:methodCdesc}
Method~C is the proposed new method of
{\em Distributed in-Cache indices}.  Unlike Method~B, the new method
intrinsically requires many nodes.
It assumes that a single node of our architecture is distinguished as
the master node, and the rest are slave nodes.  Queries always arrive
at the master node, which dispatches them to the slave nodes.

The sorted array is decomposed into equal size partitions and each 
partition is stored at a slave node in the cluster.  We assume
that each partition fits in the CPU cache.  We further assume that
there are sufficient nodes to hold these cache-sized
partitions.

\begin{figure}[hbt]
\begin{center}
  \scalebox{0.3}{\epsfig{file=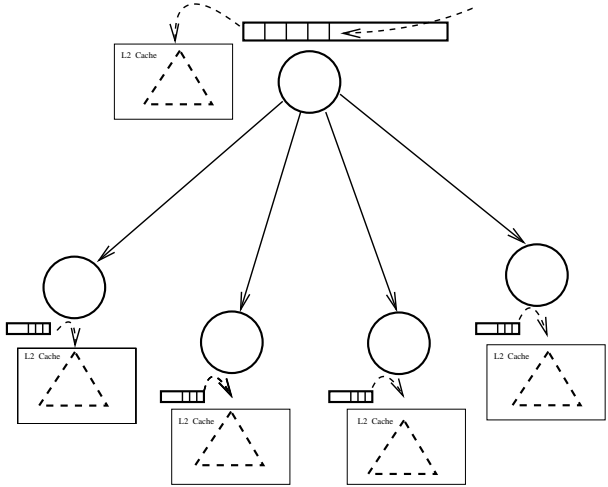}} \caption{Cooperative
  Caching Design} \label{fig:figC}
\end{center}
\end{figure}

Next, the master node contains a data structure used to determine to
which slave node the query should be dispatched.  We used a sorted
array of partition delimiters on the master node to determine to which
child a query should be passed.  This is illustrated in
Figure~\ref{fig:figC}.

The submethods~C-1, C-2 and C-3 are distinguished according to how the
slave node does the key lookup.  In method C-1, the slave node stores
its part of the index as an n-ary tree.  An optimization of Rao and
Ross~\cite{Rao00} is used to store one pointer at each node of the tree.
Given a node, its children in a tree are stored at adjacent locations. Hence,
it suffices to store only a pointer to the first child of a node.
(Rao and Ross gave this data structure the name CSB+ tree.)      

Method C-2 adds to this optimization by employing the buffered access
proposed by Zhou et al.~\cite{Zhou03}, described earlier for Method~B.  That is, the
partition on a slave node is divided into subtrees, such that each
subtree can now fit inside the L1~cache.

Method~C-3 employs a simple sorted array.  It employs binary search
for key lookup.

\medskip
\noindent
{\bf Remark.}
{\em
In principle, if there is a heavy load of incoming queries, a single
master node could become overloaded.  This is easily remedied by
setting up multiple master nodes, with replicates of the top level
data structure.
}

\section{Experimental Validation}
\label{sect:expval}
We did all experiments on a Pentium~III Linux cluster (Red Hat
release~7.2). There are 54 nodes on the Linux cluster. Each node has
two 1.3~GHz Pentium~III processors sharing 1~GB of memory.  Each
processor has its own 16~KB L1 cache and 512~KB L2 cache. The cluster
has two choices of network interconnect: a 100 Megabit/second Ethernet
and Myricom's 2.2 Gigabit/second Myrinet.  For communication, we use
the MPICH~1.2.5~\cite{mpich} implementation of MPI~\cite{mpibook}. The
default network interconnect for MPI is the 2.2 Gigabits/second Myrinet
with the GM protocol. All programs are compiled with $mpiCC$
using the $gcc-3.3.1$ compiler with
optimization level~O3.


We measured the one-way bandwidth of Myrinet as
1.1~Gb/s or 138~MB/s.  The measured memory bandwidth (Pentium~III,
266~MHz DDR RAM) was 647~MB/s for sequential memory access, and was
48~MB/s for random memory access (random access to a 4~byte word).

Note that since Method~A incurs many cache misses, the memory
bandwidth that it experiences is actually closer to the 48~MB/s quoted
above.  This is slower than the network bandwidth 138~MB/s of Myrinet,
and helps explain the experimental results.

The parameters for the tree structure used in all experiments are
reported in Table~\ref{tab:treepara} except where specifically
explained.  Both the search keys and the keys used to construct the
index structure are randomly generated.

For Methods A and B, the node size in the tree structure is equal to
the L2 cache line size. For Methods~C-1 and~C-3, the node size
in the tree structure is equal to the L1 cache line size. In
Pentium~III, both the L1 cache line size and L2 cache line size are
32~bytes. For Method~C-2, the node size is set to half size of the L1
cache to fit in the L1 cache and assistant the buffering technique. In
the implementation, the search key and the corresponding lookup result
are stored in the same memory location to lessen the cache
contention. 

\begin{table}[hbtp]
\centering
\begin{tabular}{|l|l|}
\hline Number Of Keys On The Sorted Array & 327 kilo \\ 
\hline Search Key Size & 4 bytes \\
\hline Index Tree Size & 3.2~MB \\ 
\hline Subtree Size (except the root subtree) (in B, C) & 320~KB \\ 
\hline Root Subtree Size (in B, C)& 44~bytes \\ 
\hline T (in A, B) & 7 \\ \hline L (in C-1, C-2) & 6 \\
\hline Size of Node (in A, B, and C-1) & 32~bytes \\ 
\hline Size of Root Node (in C-2) & 32~bytes \\ 
\hline Size of Leaf Node (in C-2) & 8~KB \\ 
\hline
\end{tabular}
\caption{The Index Structure Setup}
\label{tab:treepara}
\end{table}

\subsection{Comparing Methods A, B and C}
\label{sect:comparison}
In all experiments, the index tree described in
Table~\ref{tab:treepara} is applied.  We generate 8~million ($2^{23}$)
random search keys.  We use 11~nodes.  For methods A and B, the 8
million search keys are looked up locally on one node.  For method~C,
one of the 11~nodes acts as the master, and the others act as slaves.

\begin{figure*}[hbtp]
\begin{center}
\scalebox{0.5}{\epsfig{file=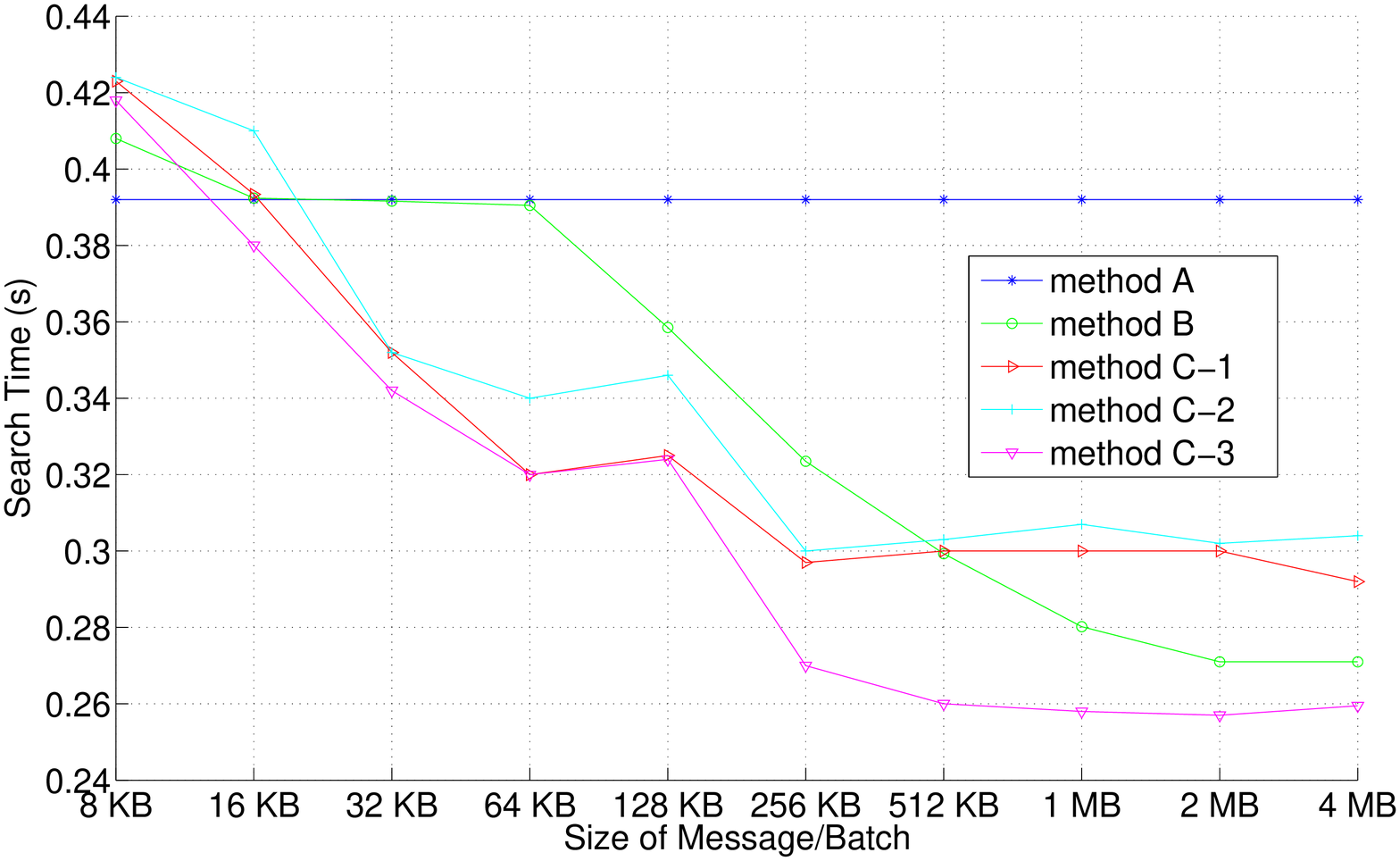}}
\end{center}
\caption{Comparing Method A, B, and C:  8~million ($2^{23}$) search
      keys (32~MB) over 11~nodes}
\label{fig:exp1}
\end{figure*}

Hence, for method~C to be competitive, it must process a search key
11~times faster than for method~A and method~B (since methods~A and~B
can use all 11~nodes, operating in parallel).  This is in fact the case,
as will be seen in Figure~\ref{fig:exp1}.  In order to make for a
fair comparison, normalization is applied to methods~A and~B: the
running time measured for a query using method~A or~B is divided
by~11.

In Figure~\ref{fig:exp1}, the x-axis shows the increase of the batch
size. The y-axis shows the running time for 8~million search keys. We
did experiments for batch sizes ranging from 8~KB to 4~MB. In
practice, one doesn't need to batch up to 4~MB. Here we just want to
show the performance trends with increasing batch sizes. In Figure~\ref{fig:exp1},
, one can see that the performance doesn't
change when the batch sizes are larger than 512~KB.  

Note also that this experimental comparison gives the benefit of doubt
to Methods~A and~B.  With 11~nodes to process queries, Methods~A and~B
require a load balancing algorithm to evenly distribute incoming
queries among all nodes.  Method~C does not require load balancing,
since all queries arrive at a single master node before being
dispatched.  In this comparison, the overhead of load balancing is
assumed to be zero.

Even after handicapping Method~C by not charging overhead for Methods~A
and~B, all of
the experiments show that Method~C-3 has the best performance.
Further, this holds according to either of two distinct measures:
throughput or response time.  The advantage of Method~C-3 with
respect to throughput is self-evident from the figure.

Figure~\ref{fig:exp1} also demonstrates the faster response time of
Method~C-3 over Method~B.  We take, as an example, the situation when
a fixed throughput of 8~million search keys in 0.32~seconds must be
processed.  The figure shows that Methods~C-2 and~C-3 achieve this
throughput with a batch size of only 64~KB, while Method~B requires a
batch size of 256~KB to achieve that same search time.  (Of course,
Method~A has a much faster response time, since it processes search
keys individually.  However, our point is that Method~C is capable of
simultaneously satisfying severe constraints in both throughput and
response time.)

Methods C-1 and C-2 follows the same trend as Method C-3 with the
increasing batch sizes, but they tend to have a slightly worse
performance.
This is because the n-ary trees of Methods~C-1 and~C-2 occupy more
space than a sorted array.  This produces more pressure on the cache.

From Figure~\ref{fig:exp1}, we see that the Methods~C are
significantly faster even for the relatively small batch sizes of
32~KB and 64~KB.  We observe a 22\% reduction in run time with this
configuration. For very large batch size, performance improvement can
still be observed even without cache coloring. If a batch size is
16~KB or less, Methods~C-1, C-2, and C-3 are worse than method B and
method A.

For a batch size of 8~KB, there are 1,000 messages, with an aggregate
communication latency of $1000\times 7$~$\mu$s.  The overhead for 8~KB
is small, and for larger batch sizes (fewer messages), the overhead is
negligible.


In the experiments, we also observed that slaves were idle for 50\% of
the time for 8~KB batch sizes, and 20\% of the time for 4~MB.  We
attribute this overhead both to the overhead of MPI and the operating
system, and statistically varying load balance among the slave nodes.
This per-message overhead is amortized across more queries as the
message size increases.  Messages were sent using MPI\_Isend in order
to overlap computation and communication to the extent supported by
the hardware.


The performance degrades slightly as the message size is increased
from 64~KB to 128~KB.  We attribute this to cache contention.  When
message sizes are 128~KB, the cache will see the 128~KB of query
lookups for the current message, 128~KB of the next message of queries
being received (overlapped communication and computation), and a
320~KB subtree for the local partition of the index.  This adds to
more than the 512~KB size of the L2~cache on the Pentium~III.

When the batch size rises beyond 128~KB, the presure of L2 cache
contention will be the same.  In that range, the benefit of the lower
slave idle time will overcome the penalty due to cache contention, and
boost the overall performance.

Our choice of 8~million search keys is for the purpose of
demonstrating the trend for larger batch sizes.  Pragmatically, one
would choose a smaller batch size for its improved response time,
while achieving similar throughput.


\subsection{Predicting the Future}
\label{sect:expmodel}

Our initial goal was to define an analytical model accurate enough to predict the
present experimental results.  For this purpose,
we wrote programs to measure the environment parameters of the
Linux cluster. We measured the memory bandwidth, network bandwidth, L2
cache line miss penalty, L1 cache line miss penalty, comparison cost
at a node whose size is equal to the L2 cache line size. These numbers
are reported in Table~\ref{tab:linuxpara}, and were used in the
analytical model (described in the Appendix).

Using the measured parameters and the equations in
Appendix~\ref{sect:analysis}, the average cost for a
query with three different methods is predicted. These are reported in
Table~\ref{tab:pretime}. We also did experiments to show the accuracy
of our evaluation.  In Table~\ref{tab:pretime}, the batch size equal
to 128~KB is applied, and one master and ten slaves are used in method
C. For fair comparison, normalization that the total running times
for Methods~A and~B are divided by~11 is applied. Table~\ref{tab:pretime} shows
that our model has over 90\% of accuracy.

\begin{table}[hbtp]
\centering
\begin{tabular}{|l|l|} \hline
Parameter & Value \\ \hline L2 Cache Size & 512~KB \\ \hline L1 Cache
Size & 16~KB \\ \hline L2 Cache line Size & 32~bytes \\ \hline L1
Cache line Size & 32~bytes \\ \hline $B_{2}\_Miss\_Penalty$ & 110~ns
\\ \hline $B_{1}\_Miss\_Penalty$ & 16.25~ns \\ \hline TLB Entries & 64
\\
\hline $Comp\_Cost\_Node$ & 30~ns \\ \hline $W_{1}$ (Memory Bandwidth) & 647~MB/s \\ \hline
$W_{2}$ (Network Bandwidth) & 138~MB/s \\ \hline
\end{tabular}
\caption{Parameters On the Linux Cluster}
\label{tab:linuxpara}
\end{table}


\begin{table}[hbtp]
\centering
\begin{tabular}{|l|l|l|l|}  
\hline Strategy & Equation & predicted    & experimental \\ 
                &          & time & time \\ 
\hline Method A: & Equation~\ref{equat:Eval_A} &
0.45~s & 0.39~s \\ 
\hline Method B: & Equation~\ref{equat:Eval_B} &
0.38~s & 0.36~s\\ 
\hline Method C-3: & Equation~\ref{equat:Eval_C} &
0.28~s & 0.32~s\\ 
\hline
\end{tabular}
\caption{Normalized Predicted and Experimental Running Time for 8 Meg ($2^{23}$)
keys}
\label{tab:pretime}
\end{table}


Table~\ref{tab:pretime} provides assurance that the model is
reasonably accurate (at least to within 25\%).  With confidence in our
present-day estimates, we go on to predict the future.

We assume that CPU speed will continue to double every 18~months,
while network speed will double only every 3~years.  Memory bandwidth
is assumed to grow, but the number of processors sharing the same memory
bandwidth may grow also with the trend to multi-processor CPU
chips. We assume the memory bandwidth available for one processor will
grow 20\% per year.  Memory latency is assumed not
to change.

\begin{figure}[hbtp]
\begin{center}
 \scalebox{0.23}{\epsfig{file=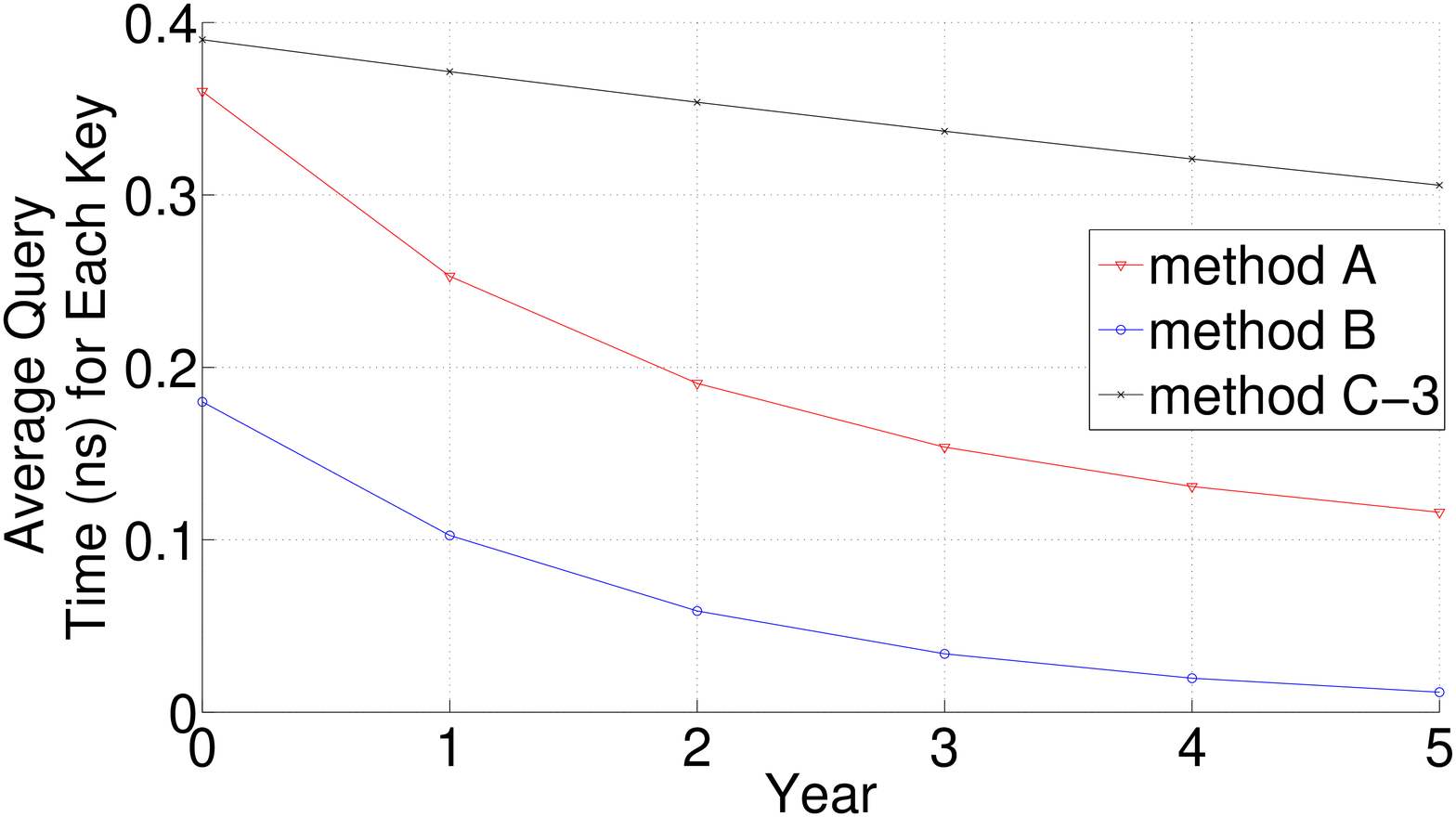}}
 \caption{Future Trends Based on Model
   (128~KB batch size, 8 meg ($2^{23}$) keys)}
\end{center}
\label{fig:future}
\end{figure}

\medskip
\noindent
{\bf Looking into the future.}
We next employ highly approximate techniques to argue the trends in
the future.  We do not claim high accuracy for this speculative
section.  However, even this crude argument will suffice to argue the
trends.

Figure~4
 demonstrates that the advantage of Method~C-3
will continue to grow.  Note that the ratio of times comparing
Method~B to~C-3 grows from approximately a factor of~2 in year~0 to
about a factor of~10 in year~5.  Methods~C-1 and~C-2 were not graphed,
for simplicity, but their curves would be close to that of C-3.

There is clearly some inaccuracy both in our analytical model as
compared to experiment, and in our assumptions of future trends.
Nevertheless, the trend of a growing speed advantage for Method~C-3 is
strong, and the conclusion is likely to remain under other scenarios
of future technological trends.

\section{Conclusion}
We proposed and evaluated the distributed in-cache indices for the tree lookup
problem. The experiments show all methods (C-1, C-2, and C-3) with distributed in-cache indices
outperform other methods when combing the two worlds, throughput and
response time. Especially, Method~C-3 is the best in most
scenario. Method~C-3 is two times faster
than Method~A and has much higher throughput and faster response time
than Method~B. Our analytical model argues that technological trends
of faster CPU and network will further favor Method~C-3.



\section{Acknowledgment}
We thank the Mariner Project at Boston University for providing the
experimental facilities.


\nocite{Ana99}  

\bibliographystyle{latex8} 
\bibliography{cooperative-caching} 

\appendix
\renewcommand{\thesection}{\Alph{section}}
\section{APPENDIX:  Analysis of Index Lookup for the  Three Methods}
\label{sect:analysis}
We introduce a model to analyze the cache
performance of a tree index structure. The model is based on the
expected number of cache line misses for each key lookup. TLB misses
are not considered in our model.  So our model gives a lower bound for
the running time. Then we apply this model to analyze three different
designs.

In our model, an n-ary tree index structure and a stream of arriving
search keys are assumed.  The variable~$n$ is chosen so that
$n$~computer words fit in an L2~cache line.

Table~\ref{tab:paramodel}, below, enumerates all the notations that will be
used in our later discussion:

\begin{table*}[hbtp]
\centering
\begin{tabular}{|l|l|}
\hline variable & Description \\

\hline $Tree\_Size$ & the size of the B+ tree \\
\hline $T$ & the total levels of the B+ tree. $T = (log(M/K)/log(K+1)
+ 1)$ \\ 
\hline $L$ & the levels of the B+ tree can fit in cache. Each
slave hold $L$ levels of the B+ tree \\ 
\hline $W_{1}$ & the memory bandwidth  647~MB/s \\ 
\hline $W_{2}$ & the network bandwidth 138~MB/s \\ 
\hline $C_{2}$ & the size of L2 cache \\ 
\hline $B_{2}\_Miss\_Penalty$ & the cost of loading a cache line from the
memory to the L2 cache \\ 
\hline $B_{2}$ & the size of the L2 cache line in bytes \\ 
\hline $B_{1}\_Miss\_Penalty$ & the cost of loading a
cache line from the L2 cache to L1 cache \\ 
\hline $B_{1}$ & the size of the L1 cache line in bytes \\ 
\hline $Comp\_Cost\_Node$ & the cost to traverse one level of the B+ tree while searching a key \\
\hline $NUM_{masters}$ & the number of master nodes \\ 
\hline $NUM_{slaves}$ & the number of slave nodes that have lower L levels of
the B+ tree in L2 cache \\
\hline $NUM_{keys\_per\_batch}$ & the number of search keys in one
batch lookup \\ 
\hline
\end{tabular}
\caption{Parameters Used in The Model}
\label{tab:paramodel}
\end{table*}

\subsection{The Model of Cache Performance for Tree Traversal}



We follow the analysis of Hankins and Patel~\cite{Hankins03}.
They assumed that the probability of accessing a vertex in a tree
depended on its level in the tree.  Hence, for an n-ary tree, the
children of the root node have probability of being accessed on the
next round that is 1/n of the probability of the root node being
accessed next.

According to~\cite{Hankins03}, for a tree that can fit in the L2
cache, the expected number of cache misses for each key lookup is:
\begin{equation}
\label{equat:Hankins03}
\frac{\sum_{i=1}^{T} X_{D}(\lambda_{i}, q)}{q}
\end{equation}
where:
\begin{equation}
X_{D}(\lambda_i, q) = \lambda_ \times (1 - (1 - 1/\lambda_i)^q) \\
\end{equation}

In the above formula, $\lambda_{i}$ is the number of cache lines at
the $i$th level of the tree, $q$ is the total number of keys to be
lookuped.

We use~\cite{Hankins03} as the foundation and further explore the
model.  We analyze the problem in two steps:

\begin{enumerate}
\item We assume that the tree space touched by the first $q_{0}$
lookups is exactly the size of L2~cache. The cache state is marked as
the state~$S0$.  The state~$S0$ represents a state when all of the
cache has become occupied by the tree structure.
\begin{equation}
\label{equat:LowMiss}
\sum_{i=1}^{T} X_{D}(\lambda_{i}, q_{0}) = C_{2}/B_{2}
\end{equation}

\item
The number of caches misses for each key lookup after the $q_{0}$th
lookup is:

\begin{eqnarray}
\label{equat:BigBatch}
\lefteqn{\sum_{i=1}^{T} X_{D}(\lambda_{i}, q_{0} + 1) - \sum_{i=1}^{T}
         X_{D}(\lambda_{i}, q_{0}) =} \\
 & & \sum_{i=1}^{T} X_{D}(\lambda_{i}, q_{0} + 1) -
 C_{2}/B_{2}\mbox{\hskip 1in}
\end{eqnarray}

For the $(q_{0}+1)$-th lookup, the amount of space loaded from
memory to cache is calculated by Equation~\ref{equat:BigBatch}. After
the $(q_{0}+1)$th lookup, the cache state is same as the state of
$S0$.  Hence, for all later lookups, each lookup needs to load
the space (calculated in Equation~\ref{equat:BigBatch}) from memory to
cache.
\end{enumerate}


\subsection{Index Structure Analysis for Search Operands}

In our evaluation, we examine only the data cache behavior, while
ignoring the instruction cache misses and TLB misses.  Instruction
cache performance is ignored, because the instruction complexity is
comparable between three methods.  Method~A and method~B are
significantly affected by TLB misses, because they work on very large
datasets.  In contrast, method C generates few TLB misses, except
immediately after a cold start.  This is because Method~C works on a small
contiguous dataset in memory.  Hence, the following analysis results
yield a lower bound running time for Methods~A and~B, but a
more accurate running time for Method~C.

\subsubsection{Method A: Standard Method} 
\label{sect:Eval_A}
For each key lookup the cost for the standard one-by-one key lookup
is:

\begin{eqnarray*}
\label{equat:Eval_A}
\lefteqn{T \times Comparison\_Cost\_Node + \frac{8}{W_{1}} +} & \\
 & & (\sum_{i=1}^{T} X_{D}(\lambda_{i},q_{0} + 1) - C_{2}/B_{2})
\times B_{2}\_Miss\_Penalty
\end{eqnarray*} 

The first term is the computation cost
and the other terms are the memory access costs. 

Any path from the
root node to one of the leaf nodes consists of $T$ nodes for a tree
with $T$ levels. So, for each search key lookup, the computation cost is $T \times
Comparison\_Cost\_Node$.

The memory access cost consists of two parts: buffer access cost and
tree access cost.  Each search key needs to be read from an
input buffer and to be written to an output buffer.  The costs of reading
from a buffer and writing to a buffer are $4/W_{1}$ each, because the input
buffer and output buffer are accessed sequentially. 

The tree access cost is calculated according to the
equation~\ref{equat:BigBatch} with $q \gg q0$. We ignore the time
spent to access data in the L2~cache, because access to data in memory
dominates the time.  Intuitively, the frequently accessed upper levels
of the tree have higher probability of remaining in cache, but the
lower levels of the tree are usually not in the cache.  Hence, a cache
miss typically happens at each level when accessing the lower parts of
the tree.



\subsubsection {Method B: The Buffering Access Method} 
\label{sect:Eval_B}

For each search key, the cost is:
\begin{eqnarray*}
\label{equat:Eval_B}
T \times Comparison\_Cost\_Node + \triangle + & \frac{4}{W_{1}} \times (T/L) + &\\
\frac{B_{2}\_Miss\_Penalty \times 4}{B_{2}} \times (T/L - 1) & &
\end{eqnarray*}

The first term represents the computation cost explained in
Section~\ref{sect:Eval_A} . The other terms represent memory access
costs. The tree access cost is $\triangle$. 

The memory access cost to read a key from buffers is $4/W_{1} \times
T/L$, because each buffer is sequentially accessed and there are a
total of $T/L$ subtrees. The total memory access cost to write a
search key to buffers is $\frac{B_{2}\_Miss\_Penalty \times 4}{B_{2}}
\times (T/L - 1)$, because each time a write buffer is
selected according to a random key value. 

The tree access cost has two parts:  the time spent to load the
subtrees from memory to L2~cache one by one ($\theta_{1}$); and the time
spent to access the subtree in the L2~cache after a subtree has been
loaded into L2~cache ($\theta_{2}$). The time spent to load all the
subtrees from memory to L2~cache can be calculated with
Equation~\ref{equat:Hankins03} because each subtree can fit in the L2
cache.  For each key lookup, the average number of L2~cache misses
are:
\begin{equation}
\theta_{1} = \frac{\sum_{i=1}^{T}X_{D}(\lambda_{i}, q)}{q} \times
B_{2}\_Miss\_Penalty
\end{equation}

For each lookup, the number of nodes to be accessed is $T$ and the
number of L2 cache lines to be accessed is also $T$, because the size
of node is same as the size of L2~cache line.  The number of cache
lines to be accessed in the L2~cache will be $T -
{\sum_{i=1}^{T}X_{D}(\lambda_{i}, q)}/{q}$. Therefore, the time
spent to access to data in the L1~cache will be:
\begin{equation}
\theta_{2} = (T - \frac{\sum_{i=1}^{T}X_{D}(\lambda_{i}, q)}{q})
\times B_{1}\_Miss\_Penalty
\end{equation}



\subsubsection{Method C: Distributed in-Cache indices for Index Structures}
\label{sect:methodC}

We make the following assumptions, which simplify the analysis.
\begin{enumerate}
\item Aggregate network bandwidth is unlimited.
\item There are enough nodes in the cluster so the the aggregate L2
  caches over the cluster can hold the entire index structure. Each
  node does computation and data accesses in cache.
\item $T < 2L$, so that each search can be done within the caches of
just two nodes: a master and a slave. Here, we make this assumption to
make the model simpler. In practice, if $T > 2L $, each search needs
to traverse more than the caches of two nodes and our design still can
be applied.
\item The master and slaves do their tasks in parallel.
\item
\end{enumerate}

For each search key, the average cost is
\begin{eqnarray*}
\label{equat:Eval_C}
 \lefteqn{\max\bigg\{\frac{Dispatch\_Cost + \frac{8}{W_{1}} + \frac{4}{W_{2}}}{num_{masters}},} \\
 & &\frac{L \times (Comp\_Cost\_Node +
      B_{1}\_Miss\_Penalty)}{num_{slaves}} \\
 & & + \frac{\frac{8}{W_{1}} +
      \frac{4}{W_{2}}} {num_{slaves}} \bigg\}
\end{eqnarray*}

In Equation~\ref{equat:Eval_C}, the first part is the cost on the
master side and the second part is the cost on the slave side. The
maximum value is the real cost because masters and slaves do tasks in
parallel. The following explains how to calculate the costs on the
master side and the slave side.

\medskip
\noindent
{\bf Cost on the master side for each search key:}
\begin{enumerate}
\item Computation time: $Dispatch\_Cost\_Per\_Search\_Key$. This cost
depends on the distribution of search key values.  We assume uniformly
distributed search key values.

\item Memory access time: $8/W_{1}$. This cost is to read a key from
the search key array and put the key to a buffer for an outgoing
message. Because accesses to the search key array and the buffer are
both sequential, the full memory bandwidth can be used to transfer data.

\item Communication time: $4/W_{2}$. For each search key, network
 transmission time is considered, but not latency.  This is because keys are
 sent out in a message with the size given of kilobyte magnitude and larger.
\end{enumerate}

The cost on the slave side for each search key:
\begin{enumerate}
\item computation time: $L \times Comparison\_Cost\_Node$. Each slave
maintains an $L$-level subtree.
\item memory access time: $8/W_{1}$. Reading a key from an incoming
message buffer and writing the result to an outgoing message buffer.
\item communication time: $4/W_{2}$. Sending the search result to the
masters. The transmission time is considered, but not latency.  This is because
results are sent in a message with the size of kilobyte magnitude or larger.
\item L2 access time: $L \times B_{1}\_Miss\_Penalty$. The tree can
fit in the L2 cache, but not in the L1 cache. For each search key, at
each level a L1 cache miss may happen.
\end{enumerate}

In Section~\ref{sect:methodCdesc}, we described three alternative
designs, {C-1}, {C-2} and {C-3}. They have similar performance.
Equation~\ref{equat:Eval_C} can be applied to all of them.

\end{document}